\documentclass[a4paper,10pt,twoside]{cpc-hepnp}
\usepackage{multicol}
\usepackage{graphicx}
\usepackage{booktabs}
\usepackage{amssymb,mathrsfs,amscd}
\usepackage[tbtags]{amsmath}
\usepackage{lastpage}
\usepackage{mathrsfs}

\begin{document}

\makeatletter
\newcommand\figurecaption{\def\@captype{figure}\caption}
\newcommand\tablecaption{\def\@captype{table}\caption}
\makeatother

\fancyhead[c]{\small Chinese Physics C~~~Vol. XX, No. X (2014)
XXXXXX} \fancyfoot[C]{\small 010201-\thepage}

\footnotetext[0]{Received 1 March 2014, Revised X X}

\title{Effect of long range correlation on the scaling behaviors of the normalized factorial moments for first-order phase transition\thanks{Supported in part by National Natural Science Foundation of China (11221504, 11075061), the Ministry of Education of China (306022) and the Programme of Introducing Talents of Discipline to Universities (B08033) }}

\author{%
      LI Guang-Lei 
\quad YANG Chun-Bin 
$^{1)}$\email{cbyang@mail.ccnu.edu.cn (corresponding author)}
}
\maketitle

\address{
 Key Laboratory of Quark and Lepton Physics (MOE) and Institute of Particle Physics, \\ Central China Normal University,
Wuhan 430079, China\\
}

\begin{abstract}
Within the framework of Ginzburg-Landau theory, the effect of multiplicity correlation between  the dynamical multiplicity fluctuations is analyzed for a first-order phase transition from quark-gluon plasma to hadrons. Normalized factorial correlators are used to study the correlated dynamical fluctuations. A scaling behavior is found among the factorial correlators, and  an approximate universal exponent, which is weakly dependent  on the details of the phase transition, is obtained.
\end{abstract}

\begin{keyword}
factorial correlators, Ginzburg-Landau model, scaling behavior
\end{keyword}

\begin{pacs}
24.85.+p, 05.70.Fh, 25.75Gz
\end{pacs}

\footnotetext[0]{\hspace*{-3mm}\raisebox{0.3ex}{$\scriptstyle\copyright$}2013
Chinese Physical Society and the Institute of High Energy Physics
of the Chinese Academy of Sciences and the Institute
of Modern Physics of the Chinese Academy of Sciences and IOP Publishing Ltd}%

\begin{multicols}{2}

\section{Introduction}

It has been known that ultra-relativistic heavy-ion collision is the only way to study the the properties of quantum chromodynamics (QCD) under extremely high energy density in the laboratory. During such a collision, a new state of matter, quark-gluon plasma (QGP) that is theoretically predicted might be formed with extremely high energy and matter density. Soon after, the system will cool with its subsequent expanding. And eventually the temperature and energy density become low enough for the hadronization process, and a phase transition may occur from QGP to hadrons.

The quarks and gluons, however, are not detectable directly in experiments because of the color confinement of QCD. We have to search for the signals about the phase transition from the final particles. Phase transition has always been a subject of great interests in high energy physics. The critical point marks the boundary of first and second order phase transition between the hadronic and QCD matter in the QCD phase diagram. The existence of a critical point has been predicted by some lattice QCD calculations \cite{lab1,lab2,lab3}. And the possibility of observing evidences for the critical point has inspired various experiments in different laboratories \cite{lab4,lab5,lab6} and a lot of relative discussions on the possible signals \cite{lab7,lab8,lab9,lab10,lab11,lab12,lab13}. So far, however, the order of the phase transition is still an open issue that has been discussed. QGP may undergo a first order or second order transition, or even a cross-over between different states with different temperature and chemical potential. Additionally, this transition may not even be recognizable as a critical phenomenon, since hadronization takes place on the surface while the system expands.

The hadrons at final state are strongly correlated and a variety of fluctuations appear. It has been known that fluctuations are large for statistical systems near their critical points, hence the study of multiplicity fluctuations of hadrons produced in high-energy heavy-ion collisions is of importance to study the phase transition \citep{
lab14,lab15,lab16,lab17,lab18}.

Ginzburg¨CLandau theory is a phenomenological model theory initially describing superconductors without examining their microscopic properties \citep{lab19,
lab20}. Over the past two decades, this model has been used to study multiplicity fluctuations about first- and second-order phase transitions \citep{lab21,lab22,lab23,lab24,lab25,lab26,lab27,
lab28,lab29,lab30,lab31,lab32,lab33,lab34}, and regarded as a possible means to reveal some features of phase transitions. Ref. \cite{lab21,lab22,lab23,
lab24,lab25,lab26,lab27,lab28,lab29,lab30} are examples to use this model to reveal dynamical multiplicity fluctuations in one bin. The QGP state, as observed experimentally, are strongly correlated according to experiments. Such correlations may influence the pattern of dynamical fluctuations for different parts
in the phase space. Therefore, the analysis of correlation to the dynamical fluctuations is of importance.
In Ref. \cite{by} the scaling behavior among the factorial correlators is studied for a second-order phase transition from QGP to hadrons. In this article, we will try to investigate the scaling behavior of the factorial correlators of multiplicity distribution within an extended Ginzburg-Landau model for a first-order phase transition for the QGP system.

This article is organized as follows. In Section 2,  the normalized  factorial correlators for multiplicity fluctuations is derived for a first-order phase transition within the framework of Ginzburg-Landau model. Section 3 is devoted to our numerical results and some conclusions. In Section 4, a concise summary is presented.

\section{Factorial correlators in the Ginzburg-Landau model for a first-order phase transition}

Consider two small bins in phase space with equal size $\delta$ (it can be an interval of a one-dimensional variable, such as rapidity $\delta y$, or that in three dimensional space, such as $\delta y\delta {\textbf{p}}_T$). Let the particle numbers in these two bins are $n_1$, $n_2$ respectively for an event. The moments of multiplicity difference has been investigated in Refs. \cite{lab31,lab32,lab33,lab34}, assuming that the fluctuations in the two bins are uncorrelated.

For the purpose of measuring the correlated fluctuations in the two bins, we can write the factorial correlators can be defined  in a similar way to Ref.  \cite{by},
\begin{align}
f_{q_1 q_2} =&\langle n_1(n_1-1)\cdots(n_1-q_1+1)
n_2(n_2-1)\cdots\nonumber\\
&(n_2-q_2+1)\rangle  \\
=&Z^{-1}\iint \mathcal{D}\phi_ 1\mathcal{D}\phi_2(\delta|\phi_1|)^{q_1}(\delta|\phi_2|)^{q_2}e^{-F(\phi_1,\phi_2)},
\end{align}
where
\begin{equation}
Z=\iint \mathcal{D}\phi_ 1\mathcal{D}\phi_2e^{-F(\phi_1,\phi_2)},
\end{equation}
where $\langle\cdots\rangle $ is the average over events,  $e^{-F(\phi_1,\phi_2)}$ is the dynamical factor for the process, and $\phi_1$, $\phi_2$ describe the probability for the systems in the two bins in pure states $|\phi_1\rangle$ and $|\phi_2\rangle$ respectively. According to the Ginzburg-Landau model, the free energy function $F(\phi_1,\phi_2)$ can be written as, for a first-order phase transition,
\begin{align}
\label{eq1}
F(\phi_1,\phi_2) =&\delta\sum_{i=1}^2(a|\phi_i|^2+b|\phi_i|^4+c|\phi_i|^6)\nonumber\\
& +\lambda\delta(|\phi_1|^2-|\phi_2|^2)^2.
\end{align}
As in Ref. \cite{lab21}, $a\propto(T-T_c)$ and it is negative for the hadron phase, whereas $c$ is positive. And $b$ is negative for the first-order transition.
The last term $\lambda\delta(|\phi_1|^2-|\phi_2|^2)^2$ in Eq.~(\ref{eq1}) is introduced to parameterize the effect of correlation between particle productions in the two bins. The parameter $\lambda$ is used to describe the strength of interactions in the two bins. If $\lambda>0$, the free energy $F$ is smaller for $|\phi_1|$ is closer to
$|\phi_2|$ and the particle production in the two bins is positively correlated. Otherwise, particle production
in the two bins is anti-correlated for $\lambda<0$. The absolute value of $\lambda$ should decreases as the distance between the two bins in a phase space increases. Furthermore, as the correlation length of the QGP system is longer near the critical point, $\lambda$ may also have a relation with the temperature departure from the critical point. Then its value can also mirror the degree of separation from critical temperature, if the distance between the two bins in phase space is fixed. This paper is confined only to $\lambda>0$ but the extension to $\lambda<0$ is obvious.

The normalized correlated factorial moments can be defined as
\begin{equation}
F_{q_1 q_2}=f_{q_1 q_2}/[(f_{1,0})^{q_1}(f_{0,1})^{q_2}].
\end{equation}
\noindent
Since $F_{q_1 q_2}$ will be constants of about 1 if there are only statistical fluctuations, the moments can be used to filter the statistical fluctuations. The so-called intermittency behavior is for a phenomenon in which
$F_{q_1q_2}\propto \delta^{-\alpha_{q_1q_2}}$ with $\alpha_{q_1q_2}>0$. What is more, one can further study
whether there exists a scaling law among $F_{q_1q_2}$,
\begin{equation}
F_{q_1 q_2}\propto F_{22}^{\beta_{q_1q_2}}\ ,\label{scal}
\end{equation}
\noindent even when the intermittency behavior can not be observed.
If there exists no correlation between multiplicity fluctuation in the two bins, i.e. $\lambda=0$, then the factorial moments are simply
\begin{equation}
\label{eq2}
F_{q_1 q_2}=F_{q_1}F_{q_2},
\end{equation}
with $F_q$ being the normalized factorial moments for multiplicity fluctuations in  one bin. Then the scaling
  behavior among $F_{q_1q_2}$ is the same as among $F_q$. If there exists correlation between the multiplicity fluctuations in the two bins, factorization shown by Eq.~(\ref{eq2})
is not valid, then whether the scaling behaviors in Eq. (\ref{scal}) are still valid is a problem that needed to be solved.

By defining
\begin{equation}
\label{eq3}
J_q(z_1,z_2)=\int_0^\infty  dy y^{q} e^{-y^3+z_1y+z_2y^2}
\end{equation}
the factorial normalized moments can be finally written as
\begin{align}
F_{q_1 q_2}=&\frac{\int_0^\infty dx x^{q_1} e^{-x^3+wx+vx^2}J_{q_2}(u,v)}{[\int_0^\infty dx x e^{-x^3+wx+vx^2}J_0(u,v)]^{q_1}}\cdot \nonumber\\
&\frac{[\int_0^\infty dx e^{-x^3+wx+vx^2}J_0(u,v)]^{q_1+q_2-1}}{[\int_0^\infty dx e^{-x^3+wx+vx^2}J_1(u,v)]^{q_2}},
\end{align}
where $u=w+\sqrt{w}sx,v=\sqrt{w}t$, $s=2\lambda/\sqrt{|ac|}$, $t=-(b+\lambda)/\sqrt{|ac|}$, and
 $w=-a\frac{\delta^{\frac{2}{3}}}{c^{\frac{1}{3}}}$. Thus
$w$ can be regarded as a measure of the bin size.

Since only the first-order transition is considered in this paper, $a<0$, $b<0$, $c>0$. If one supposes $\lambda>0$, then $w>0$, $s>0$, but $t$ may be negative or positive.
From the definition of $J_q$ in Eq.~(\ref{eq3}) , we can get iterative relations as follows,
\begin{align}
J_2(z_1,z_2)&=\frac{1}{3}+\frac{1}{3}(z_1J_0(z_1,z_2)+2z_2J_1(z_1,z_2)),\nonumber\\
J_q(z_1,z_2)&=\frac{1}{3}[(q-2)J_{q-3}(z_1,z_2)+z_1J_{q-2}(z_1,z_2)+2z_2J_{q-1}(z_1,z_2)].\nonumber
\end{align}
For simplicity only the case for $q_1=q_2$ is discuss, and a more general case is left for later study.

\section{Numerical results and discussions}
From the above relations, we can calculate the normalized factorial correlators $F_{qq}$ as a function of $w$ (or the bin size resolution $\delta$) with pre-specified $s$ and $t$.

In order to get the numerical results about factorial correlators, we first fix  the parameters $s$ and $t$ both equal to 0.2 and analyse the dependence of $F_{qq}$ as functions of $-\ln w$ on the bin size, while q is from 2 to 7. The results are shown in Fig.~\ref{fig1}. As displayed in this figure, with the decrease of bin size $\delta$ (or parameter $w$), i.e. the increase of $-\ln w$ in the figure, $F_{qq}$ increase monotonically. This can be explained as follows. For larger bins, different dynamical fluctuations perhaps counteract each other, which renders them less observable.

\begin{center}
\includegraphics[scale=0.5]{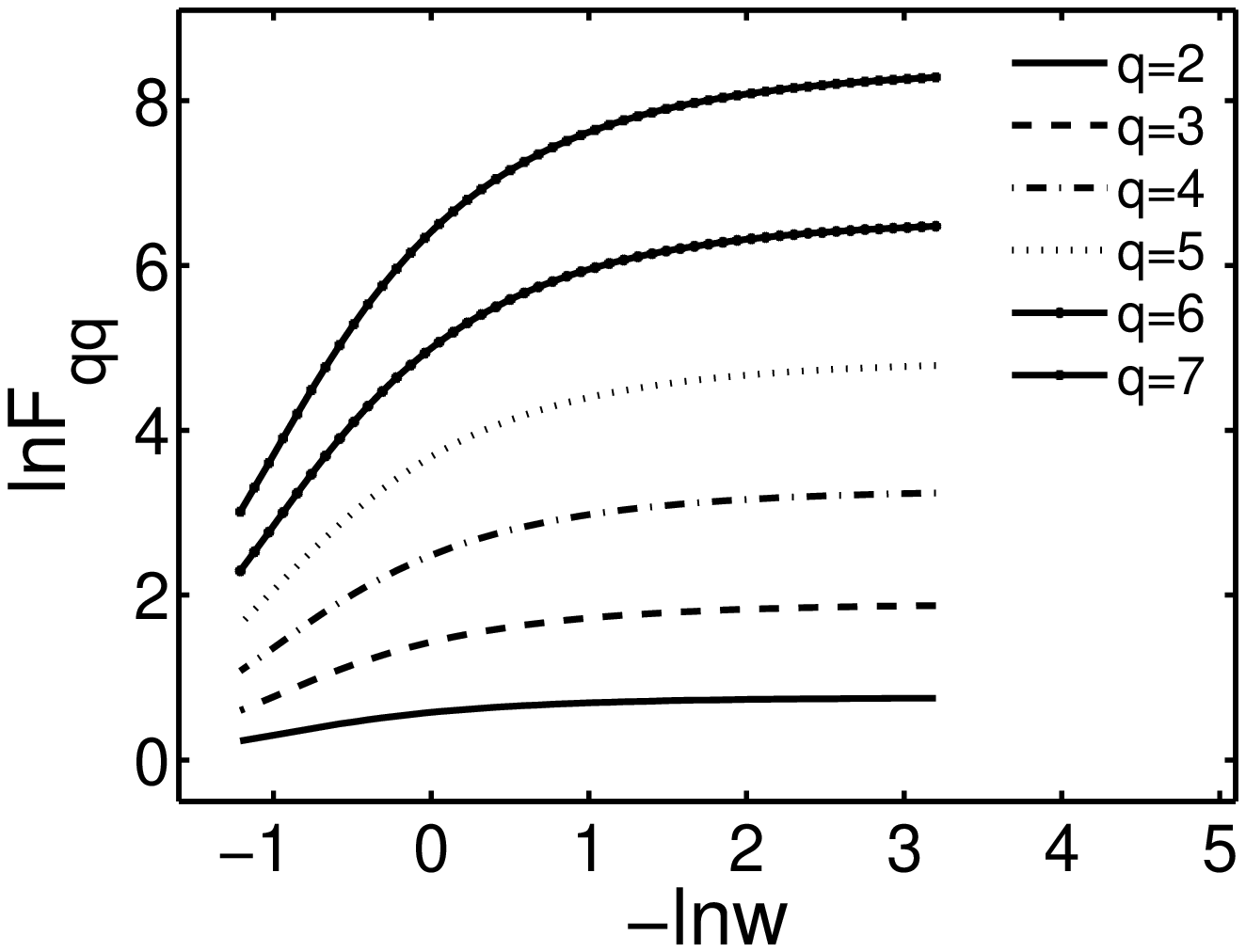}
\figcaption{\label{fig1} The dependence of $F_{qq}$ on bin size (represented by $w$), for $q$ from 2 to 7, with parameter $s$ and $t$ fixed at 0.2. }
\end{center}

 As discussed in Ref. \cite{lab21}, for a self-similar dynamical process, the moments $F_{qq}$ will be a power law function of bin size $\delta$ or parameter $w$, i.e. $F_{qq}\propto w^{-\phi_q}$. It is obvious that intermittency behavior is not observed from Fig.~\ref{fig1}, since the curves are not linear for the log-log coordinate.

The similar behaviours among $F_{qq}$ in Fig.~\ref{fig1} imply a quite simple relation among $F_{22}$ and $F_{qq}$. A power-law among $F_{22}$ and $F_{qq}$ for different values of $q$ can be reached
\begin{equation}
\label{eq4}
F_{qq}\propto F_{22}^{\beta_q},
\end{equation}
which is more general, for Eq.~(\ref{eq4})  can still hold even if the law of intermittency is violated. The relation displayed in Eq.~(\ref{eq4}), namely the scaling behavior, can be observed in Fig.~\ref{fig2} for $s=0.2$, $t=0.2$ and $q$ from 3 to 7, since all the curves in the figure can be well approximated by linear lines.

The exponent $\beta_q$ is dependent on q, parameters s and t. To find an exponent that is independent of details of our model, we can present  $\beta_q$ as a function of $q-1$. The result is shown in Fig.~\ref{fig3} in
 log-log scale. Additionally, we also plot a linear fit in this picture, and immediately get
\begin{equation}
\label{eq5}
\beta_q\propto(q-1)^{\gamma}
\end{equation}
with $\gamma=1.293$, which depends only on the values of parameters $s=0.2$, $t=0.2$

\begin{center}
\includegraphics[scale=0.5]{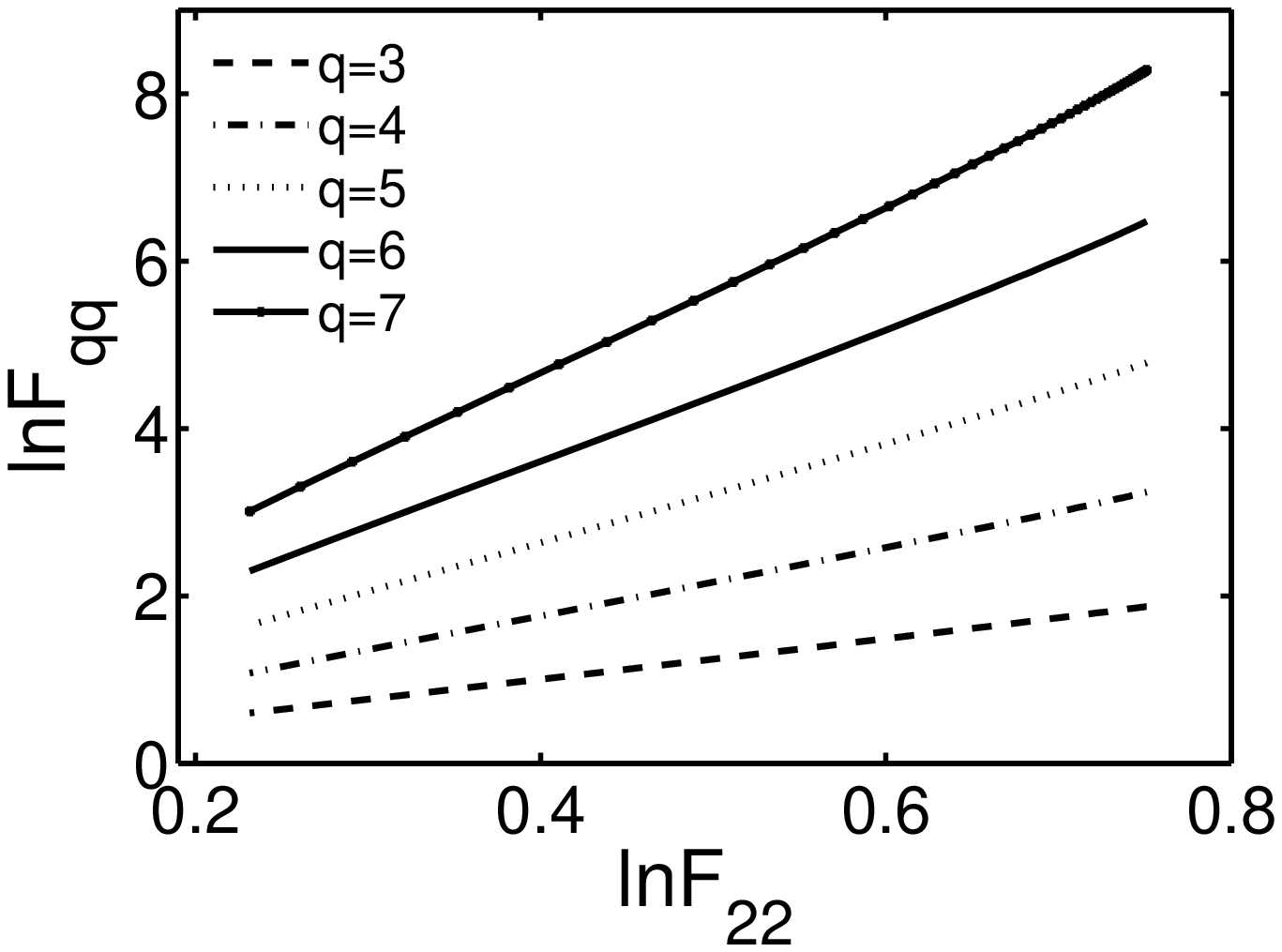}
\figcaption{\label{fig2}  Scaling behaviour between $F_{qq}$ and $F_{22}$, for q from 3 to 7, and $s$ and $t$ fixed at 0.2.}
\end{center}

\begin{center}
\includegraphics[scale=0.5]{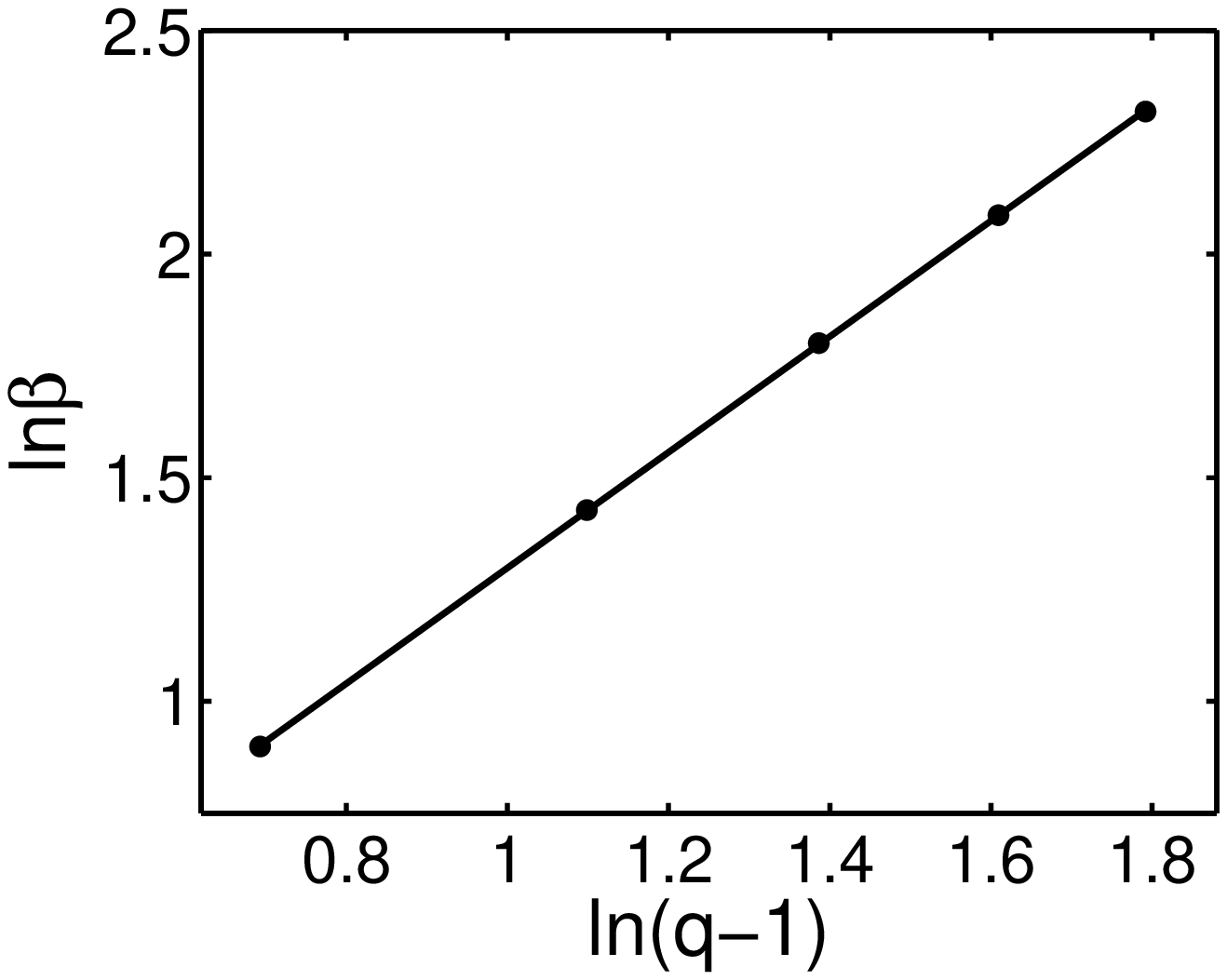}
\figcaption{\label{fig3}  Relation between $\beta_q$ and $q$, for $q$ from 3 to 7 with $s$ and $t$ fixed at 0.2. This relation can fit the scaling behavior $\beta_q\propto(q-1)^{\gamma}$,  with $\gamma=1.297$.}
\end{center}

Next task is to study whether the same scaling behavior can be found for other values of $s$ and $t$. We have analysed some other different values for $s$ and $t$, and got the corresponding $\beta_q$ and $\gamma$ just like the procedures showed above. And we finally come to the conclusion that the scaling relation, $F_{qq}\propto F_{22}^{\beta_q}$ is still valid for other values of $s$ and $t$ also.

As is depicted in Fig.~\ref{fig4} and Fig.~\ref{fig5}, we can also find the exponent value $\gamma$ of Eq.~(\ref{eq5}) depends weekly on parameter $s$ in the range from $0.2$ to $1.0$ and parameter $t$ between $-0.4$ and $1.0$. $\gamma$ is about $1.29\pm0.01$  for other $s$ and $t$ in these two figures. Comparing to the case of no interaction, i.e. $s=0$, it can be clearly seen that the exponent $\gamma$ is very weekly dependent on the details of interaction and only sensitive to the phase transition. Consequently, it can be regarded as a well observable quantity to characterize the nature of phase transition.

\begin{center}
\includegraphics[scale=0.5]{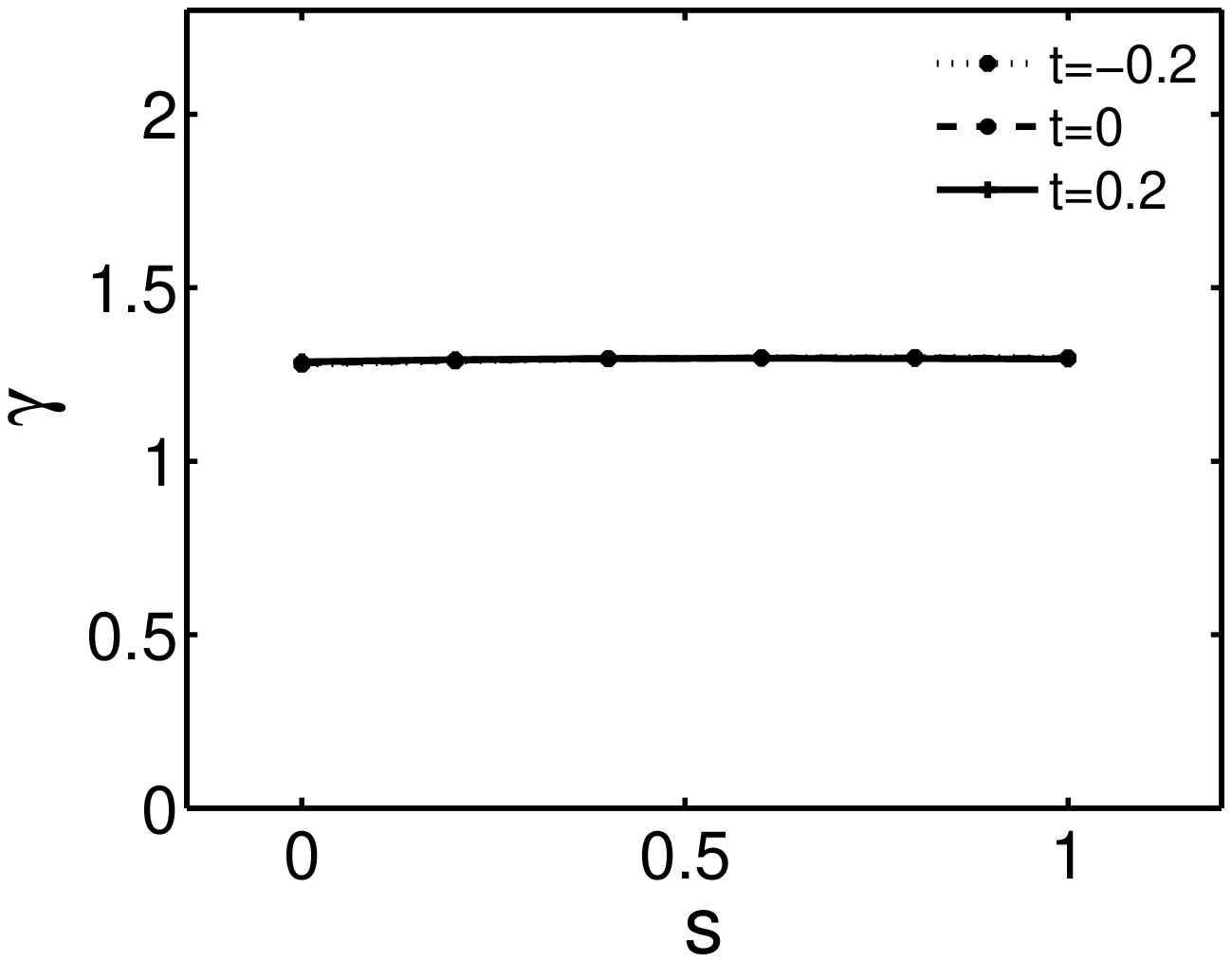}
\figcaption{\label{fig4}  Dependence of $\gamma$ on $s$ with parameter $t$  fixed at three different values $-0.2$, $0$ and $0.2$.}
\end{center}

\begin{center}
\includegraphics[scale=0.5]{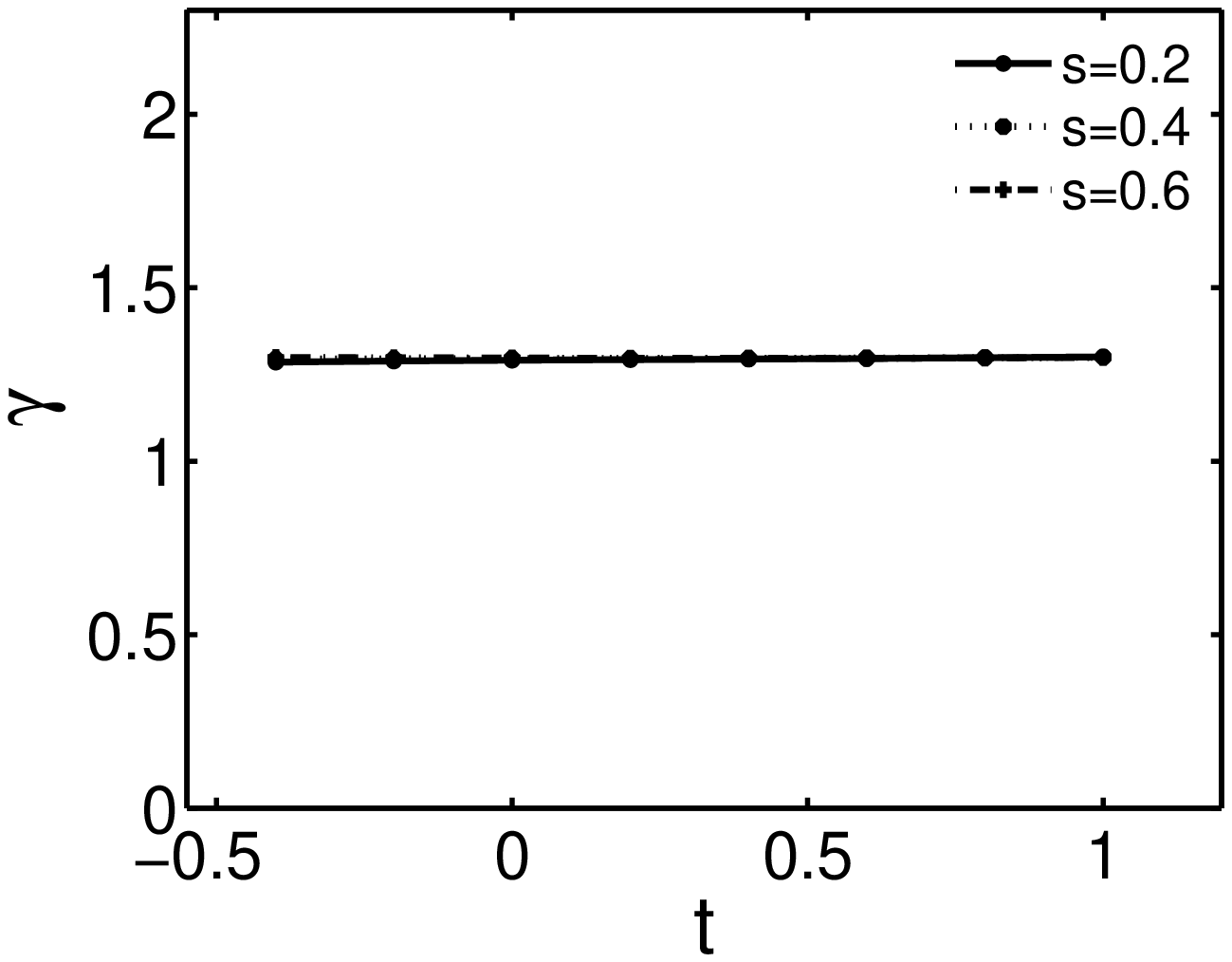}
\figcaption{\label{fig5}  Dependence of$\gamma$ on $t$ withe parameter $s$  fixed at $0.2$, $0.4$ and $0.6$.}
\end{center}

\section{Summery}
We have studied the scaling behavior of the normalized factorial correlators for correlated multiplicity fluctuations in a first-order transition from QGP to hadrons, and found a universal scaling exponent $\gamma=1.29\pm0.1$, which is nearly independent of dimension of phase space and the details of interactions. Therefore, it provides a practical quantity that can characterize the dynamical fluctuations during the phase transition.



\vspace{-1mm}
\centerline{\rule{80mm}{0.1pt}}
\vspace{2mm}


\end{multicols}

\clearpage


\begin{thebibliography}{90}

\vspace{3mm}

\bibitem{lab1} Fodor Z, Katz S D, J. High Energy Phys., 2004, {\bf 04}: 050
\bibitem{lab2} Ejiri S. Phys. Rev. D, 2008, {\bf 78}: 074507
\bibitem{lab3} Gavai R V, Gupta S. Phys. Rev. D, 2008, {\bf 78}: 114503
\bibitem{lab4} Hidaka Y, Yamamoto N. Journal of Physics: Conference Series, 2013, {\bf 432}:012017

\bibitem{lab5} Mohanty B. Nucl. Phys. A, 2009, {\bf 830}: 899-907
\bibitem{lab6} Mohanty B. J. Phys. G: Nucl. Part. Phys., 2011, {\bf 38}: 124023
\bibitem{lab7} Lacey R A, Ajitanand N N, Alexander J M et al. Phys. Rev. Lett., 2007, {\bf 98}: 092301

\bibitem{lab8} Hatta Y, Stephanov M A. Phys. Rev. Lett., 2003, {\bf 91}: 102003
\bibitem{lab9} Stephanov M, Rajagopal K, Shuryak E. Phys. Rev. Lett., 1998, {\bf 81}: 4816-4819
\bibitem{lab10} Hatta Y, Ikeda  T. Phys. Rev. D, 2003,  {\bf 67}: 014028
\bibitem{lab11} Suleymanov M K, Khan E U, Ahmed K et al. Indian J. Phys., 2011, {\bf 85}: 1047-1050
\bibitem{lab12} Bass S A, Petersen H, Quammen C et al. Cent. Eur. J. Phys., 2012, {\bf 10}: 1278-1281
\bibitem{lab13} Nahrgang M, Schuster T, Mitrovski M et al. J. Phys. G: Nucl. Part. Phys., 2011, {\bf 38}: 124150


\bibitem{lab14} Tarnowsky T J. J. Phys. G: Nucl. Part. Phys., 2011, {\bf 38}: 124054

\bibitem{lab15}Hwa R C, Yang C B. Phys. Rev. C, 2012, {\bf 85}: 044914

\bibitem{lab16} Friman B, Karsch F, Redlich K et al. Eur. Phys. J. C, 2011, {\bf 71}: 1694

\bibitem{lab17} Gorenstein M I. Phys. Part. Nucl., 2008, {\bf 39}: 1102-1109

\bibitem{lab18} Konchakovski V P, Gorenstein M I, Bratkovskaya E L. Indian J. Phys., 2011, {\bf 85}: 1-4


\bibitem{lab19} Ginzburg V L, Landau L D. Zh. Eksp. Teor. Fiz., 1950, {\bf 20}: 1064
\bibitem{lab20} Gor'kov L P. Sov. Phys. JETP, 1959, {\bf 36}: 1364



\bibitem{lab21} Hwa R C. Phys. Rev. D, 1993, {\bf 47}: 2773-2781  
\bibitem{lab22} Lebedev I A, Nazirov M T. Mod. Phys. Lett. A, 1994, {\bf9}:2999  
\bibitem{lab23} Hwa R C. Phys. Rev. C, 1994, {\bf 50}: 383
\bibitem{lab24} Mohanty A K, Kataria S K. Phys. Rev. Lett., 1994 ,{\bf73}: 2672
\bibitem{lab25} Cai X, Yang C B, Zhou Z M. Phys. Rev. C, 1996, {\bf 54}: 2775-2778   
\bibitem{lab26} Yang C B, Cai X. Phys. Rev. C, 1998, {\bf57}: 2049 
\bibitem{lab27} Yang C B, Cai X. Phys. Rev. C, 1998, {\bf58}: 1183 
\bibitem{lab28} Yang C B, Cai X. J. Phys. G, 1999, {\bf25}: 485  
\bibitem{lab29} Yang C B, Cai X. Phys. Rev. C, 2000, {\bf61}: 014902 
\bibitem{lab30} Babichev L F, Klenitsky D V, Kuvshinov V I. Phys. Lett. B, 1995, {\bf345}: 269-271 

\bibitem{lab31} Yang C B, Cai X. J. Phys. G: Nucl. Part. Phys., 1998, {\bf 24}: 1957-1963
\bibitem{lab32} Hwa R C. Phys. Rev. D, 1998, {\bf 57}: 1831
\bibitem{lab33} Yang C B, Cai X. Phys. Rev. C, 1998, {\bf 57}: 2049
\bibitem{lab34} Yan W B, Yang C B, Cai X. Chin. Phys. Lett., 1999, {\bf 16}: 253
\bibitem{by} Bai X Z, Yang C B, Int. J. Mod. Phys., 2013,  {\bf E 22}: 1350059
\end{thebibliography}
\end{document}